\documentclass[preprint,amsmath,amssymb,floatfix,showpacs]{revtex4}

\usepackage{graphicx}
\usepackage{bm}

\begin{document}

\title{Efficient models for micro-swimmers}

\author{Nobuhiko Watari}
 \email{nobuhiko@umich.edu}
\affiliation{
Macromolecular Science and Engineering Center, University of Michigan, Ann Arbor, Michigan  48109-2136, USA
}

\author{Ronald G. Larson}
\affiliation{
Department of Chemical Engineering, University of Michigan, Ann Arbor, Michigan 48109-2136, USA
}%

\date{\today}

\begin{abstract}
We propose minimal models of one-, two- and three-dimensional micro-swimmers at low Reynolds number with a periodic non-reciprocal motion.
These swimmers are either ``pushers'' or ``pullers'' of fluid along the swimming axis, or combination of the two, depending on the history of the swimming motion. 
We show this with a linear three-bead swimmer by analytically evaluating the migration speed and the strength of the dipolar flow induced by its swimming motion. 
It is found that the distance traveled per cycle and the dipolar flow can be obtained from an integral over the area enclosed by the trajectory of the cycle projected onto a cross-plot of the two distances between beads.
Two- and three-dimensional model swimmers can tumble by breaking symmetry of the swimming motion with respect to the swimming axis, as occurs in the tumbling motion of \textit{Escherichia coli} or \textit{Chlamydomonas}, which desynchronize the motions of their flagella to reorient the swimming direction.
We also propose a five-bead model of a ``corkscrew swimmer'', i.e. with a helical flagellum and a rotary motor attached to the cell body.
Our five-bead swimmer is attracted to a nearby wall, where it swims clockwise as observed in experiments with bacteria with helical flagella.
\end{abstract}

\maketitle

\section{Introduction}

Micro-swimmers, including motile bacteria and artificial swimmers, swim at low Reynolds number by using a periodic non-reciprocal motion, i.e. motion lacking time-inversion symmetry.
Since their motion is governed by the Stokes equation, which is linear, a reciprocal motion does not induce any time-average net migration of the center-of-mass~\cite{Purcell}.
Micro-swimmers break reciprocality of the motion, for example, by rotating one or more helical flagella attached to the cell body that counter-rotates as a result (e.g. \textit{E. coli}), or by performing ``power'' and ``recovery'' strokes or the breaststroke motion with flagella (e.g. \textit{Chlamydomonas}), or by propagating a wave on a flexible flagellum from the base of the cell body to the free end (e.g. sea urchin spermatozoon)~\cite{CellMovement}.
These micro-swimmers can be categorized into ``pullers'', ``pushers'' or combinations of the two, depending on the swimming motion.
We here define a puller/pusher as a micro-swimmer that on average pulls/pushes fluid to/away from the cell along the swimming axis over one period of its swimming motion.
Examples of pullers are \textit{Chlamydomonas} and \textit{Chrysomonad}, and pushers include the sea urchin spermatozoon and \textit{E. coli}.
Some micro-swimmers, such as \textit{E. coli} and \textit{Chlamydomonas}, intersperse straight ``running motion'' with ``tumbles'' allowing them to reorient their movement stochastically towards a more favorable environment or possibly to reduce their chances of encountering predators~\cite{Berg-book,Polin,Tumbling-Stealth}

The two-bead swimmer has been proposed as a simple model of a micro-swimmer to study the collective dynamics of the swimmers and their behavior near a wall~\cite{2-bead-model,Underhill}.
However, since the two-bead swimmer does not account for a periodic non-reciprocal swimming motion, it can not generate the time-dependent flow around the swimmer, which is stronger than the time-averaged flow~\cite{Bead-spring-Ecoli} and which  can induce synchronization of the swimming motions of multiple swimmers.
Thus, the three-bead swimmer, which is the simplest model that accounts for a periodic non-reciprocal motion, has also been proposed~\cite{Najafi} and used to study the interactions of multiple swimmers~\cite{3-bead-model}.
However, as we shall describe here, the swimming motion in these early studies on the three-bead swimmer is actually a combination of simpler puller and pusher swimming motions.

Here, we propose minimal one-, two- and three-dimensional puller/pusher micro-swimmer models, and two- and three-dimensional run-and-tumble swimming motions.
By simulations using a bead-spring model, we find that the tumbling motion can be easily induced by breaking symmetry of the swimming motion with respect to the swimming axis.

Finally, we propose a five-bead swimmer as a minimal model for a swimmer with a helical flagellum and rotary motor attached to the cell body (i.e. ``corkscrew swimmer''~\cite{Purcell}).
This five-bead swimmer accounts for the hydrodynamic effects of the rotation of the flagellum and the counter-rotation of the cell body as well as the periodic non-reciprocal swimming motion.
It is found that this swimmer, when placed near a wall, reproduces the behavior of \textit{E. coli}, which is attracted to the wall~\cite{Berke} and swims clockwise, when viewed perpendicular to the wall through the fluid~\cite{Frymier}, despite the fact that the swimmer does not have a helical tail nor helical motion.

\section{Simulation method}
\begin{figure}
\begin{center}
\centerline{\includegraphics[width=5in]{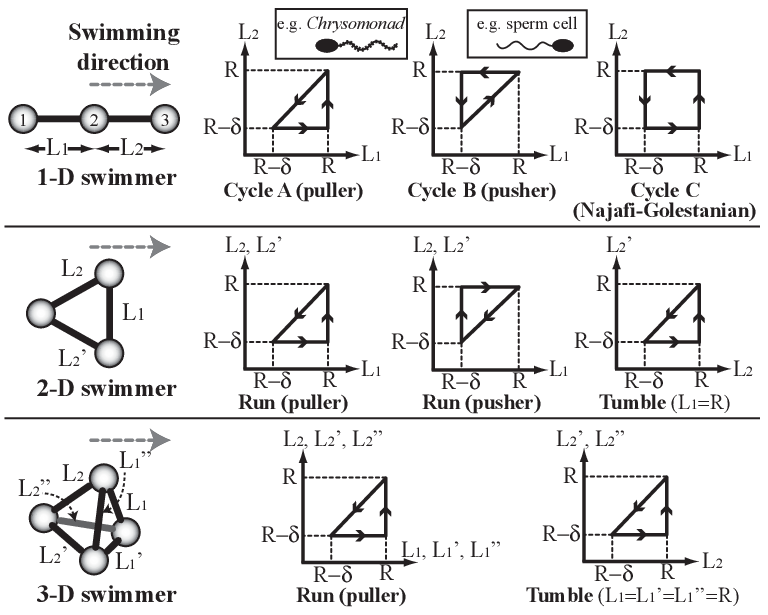}}
\caption{Minimal models of one-, two- and three-dimensional micro-swimmers and the {cycle} histories. \textit{Chrysomonad} and sea urchin spermatozoon are examples of one-dimensional puller and pusher, respectively.
Note that tumbling is induced in the 2D- and 3D-swimmers by holding the equilibrium lengths of one (2D) or three (3D) bond-lengths fixed while varying the lengths of the other bonds.
}
\label{Swimmers}
\end{center}
\end{figure}

Our minimal model of a one-dimensional swimmer consists of three colinear beads  connected by two bonds whose lengths change in a non-reciprocal manner. This geometry is identical to that of previous studies~\cite{Najafi,3-bead-model}, but we induce different cycles (or periodic histories) of bond-lengths.
Examples of non-reciprocal cycles of the two bond lengths, $L_{1}$ and $L_{2}$, that we impose are depicted in $L_{1}$-$L_{2}$ space in Fig.~\ref{Swimmers}.
This swimmer can swim at low Reynolds number by repeating a non-reciprocal cycle of bond-lengths.

We confirmed that the bond-length histories in Fig.~\ref{Swimmers} induce migration of the swimmer by simulations using the bead-spring model.
In the simulations, three spherical beads with hydrodynamic radius $a$ are connected by two FENE-Fraenkel springs~\cite{FFspring} of time-dependent equilibrium lengths $L_{1}(t)$ and $L_{2}(t)$,  which follow a configurational history shown in Fig.~\ref{Swimmers}.
The FENE-Fraenkel spring force $\bm{F}_\text{FF}$ with equilibrium length $L$ is written as:
\begin{alignat}{4}
	\bm{F}_\text{FF}(\bm{r}_{ij})
		&= H~\frac{r_{ij}/L-1}{1-\left( 1-r_{ij}/L \right)^{2}/s^{2}}\frac{\bm{r}_{ij}}{r_{ij}}\\
		&\hspace{12pt} \text{for}\hspace{2pt} (1-s)< r_{ij}/L< (1+s),\nonumber
\end{alignat}
where $i$ and $j$ are indexes of beads, $\bm{r}_{ij}$ is a vector connecting $\bm{r}_j$ (position vector of bead $j$) to $\bm{r}_i$,  $H$ is the spring constant, and the deformed spring length $r_{ij}$ is restricted to a range set by the parameter $s$, which is $0.01$ in our simulations.
The motion of each bead of the swimmer is computed according to the following equation with the force distribution on beads given by the spring forces,
\begin{alignat}{4}
\label{EqOfMotion}	& \bm{v}_{i}(t)= \sum_{j=1}^{N} {\mathcal{H}_{ij}(\bm{r}_{ij})\cdot \bm{f}_{j}},\\
	\text{or}\hspace{12pt} &\bm{r}_{i}(t+\Delta t)=\bm{r}_{i}(t)+
							\left\{
								\sum_{j=1}^{N} {\mathcal{H}_{ij}(\bm{r}_{ij})\cdot \bm{f}_{j}}
							\right\} \Delta t,
\end{alignat}
where $\bm{v}_{i}(t)$ is the velocity vector of bead $i$ at time $t$,
$\bm{r}_{i}$ is the position vector of bead $i$,
$\bm{f}_{j}$ is the force on bead $j$,
$N$ is the total number of beads in the swimmer,
$\Delta t$ is the time increment in the simulation,
and $\mathcal{H}_{ij}$ is a hydrodynamic interaction tensor.
For $\mathcal{H}_{ij}$, we use the Oseen tensor~\cite{Doi-Edwards} with {the point-force approximation}, or the Rotne-Prager-Yamakawa (RPY) tensor~\cite{RP,Yamakawa} with more accurate hydrodynamics up to terms of order $r_{ij}^{-3}$.

\begin{figure}
\begin{center}
\centerline{\includegraphics[width=2in]{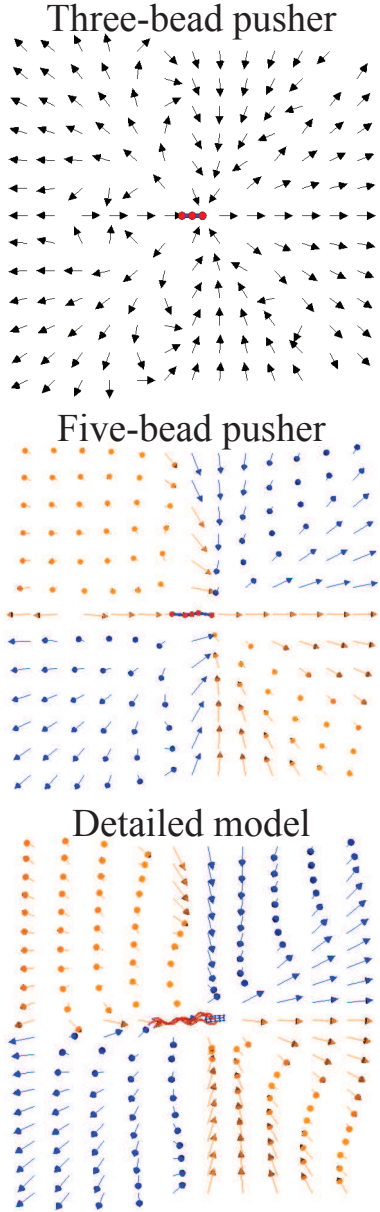}}
\caption{(color online). The time-averaged flow field around a three-bead pusher (top), around a five-bead pusher (middle) and around a detailed bead-spring model of \textit{E. coli} in a run~\cite{Bead-spring-Ecoli} (bottom).
All arrows are unit flow-velocity vectors.
Red (or light gray) arrows point into the paper, blue (or dark gray) arrows the opposite, and black arrows only have in-plane velocities. 
Swimming is from left to right.
}
\label{Flows}
\end{center}
\end{figure}

By superposing the hydrodynamic interactions between beads,
this equation with the spring forces on beads captures the dynamics of the swimmer.
Although both cycles A and B in Fig.~\ref{Swimmers} induce migration of the center-of-mass at the same speed in the same direction, our simulations show that the resulting cycle-averaged flow fields  differ.
The far-field cycle-averaged flow of cycle A  corresponds to that of a puller and cycle B to a pusher, and both flows decay as $r^{-2}$.
The cycle-averaged flow of the pusher is surprisingly similar to that of a detailed bead-spring model of \textit{E. coli} in a run~\cite{Bead-spring-Ecoli} except that the flow of the three-bead swimmer does not have angular velocity components around the swimming axis (see Fig.~\ref{Flows}).

By combining the {cycle} histories of the pusher and the puller, we can create a ``square'' history in $L_{1}$-$L_{2}$ space shown in Fig.~\ref{Swimmers}:cycle C, which corresponds to the swimming motion of the Najafi-Golestanian swimmer~\cite{Najafi}.
The cycle-averaged flow of this history is a combination of the flows of a puller and a pusher, which decays as $r^{-3}$ because the dipole components of the flow cancel in a cycle that consists of puller and pusher swimming motions.
It is also possible to generate a net puller or a pusher by setting different ranges for the changes in $L_{1}$ and $L_{2}$~\cite{3-bead-model} (i.e. by making the ``square'' history into a ``rectangular'' one).

\section{Migration speed of three-bead swimmer}
Here, we  evaluate the migration per cycle of a three-bead swimmer whose cycle is represented by an arbitrary closed  path in $L_{1}$-$L_{2}$ space.
Once a cycle motion is specified in $L_{1}$-$L_{2}$ space, the corresponding spring-force cycle in $f_{1}^{s}$-$f_{2}^{s}$ space is also specified by Eq.~\ref{EqOfMotion}, where $f_{i}^{s}$ is the spring force connecting beads $i$ and $i+1$, $\dot{L}_{i}$ is the time-derivative of $L_{i}$, and $\eta$ is the solvent viscosity:
\begin{equation}
\label{force-history}
	f_{i}^{s}=f_{i}^{s}(L_{1},L_{2},\dot{L}_{1},\dot{L}_{2},a,\eta).
\end{equation}
Therefore, we can calculate the migration per cycle $L_{c}$ requiring a cycle time $T$
based on the histories of $(L_{1},L_{2})$ and $(f_{1}^{s},f_{2}^{s})$ as,
\begin{equation}
\label{def-migration}
	L_{c}=\int_{0}^{T}\frac{1}{3}\sum_{i=1}^{3}{v}_{i}(t)dt.
\end{equation}
The migration per cycle does not depend
on the cycle time $T$ because of the linearity of the Stokes equation, and only depends on the cycle path in $L_{1}$-$L_{2}$ space (i.e. $L_{c}$ is an invariant of a cycle).
For example, if the time spent for a portion of a cycle becomes shorter (e.g. $\times 0.5$) following the same path, the migration speed becomes faster (e.g. $\times 2$).
As a result, the migration during the portion of the cycle stays the same.

\begin{figure}
\begin{center}
\centerline{\includegraphics[width=5in]{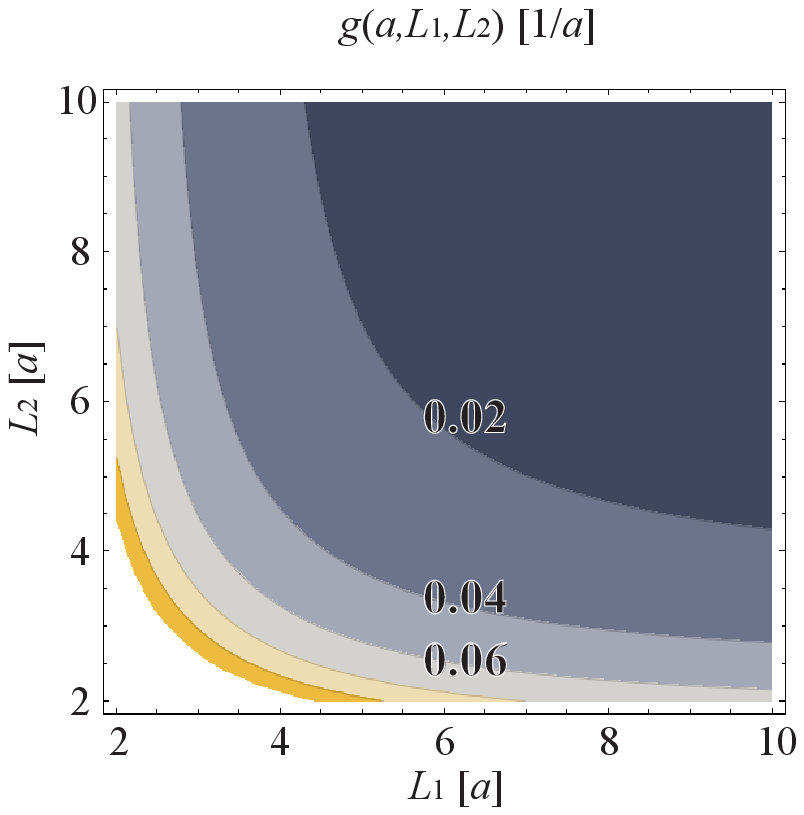}}
\caption{(color online). Contour plot of function $g(a,L_{1},L_{2})$ in Eq.~\ref{migration}.
}
\label{Fig-g}
\end{center}
\end{figure}

\begin{figure}
\begin{center}
\centerline{\includegraphics[width=5in]{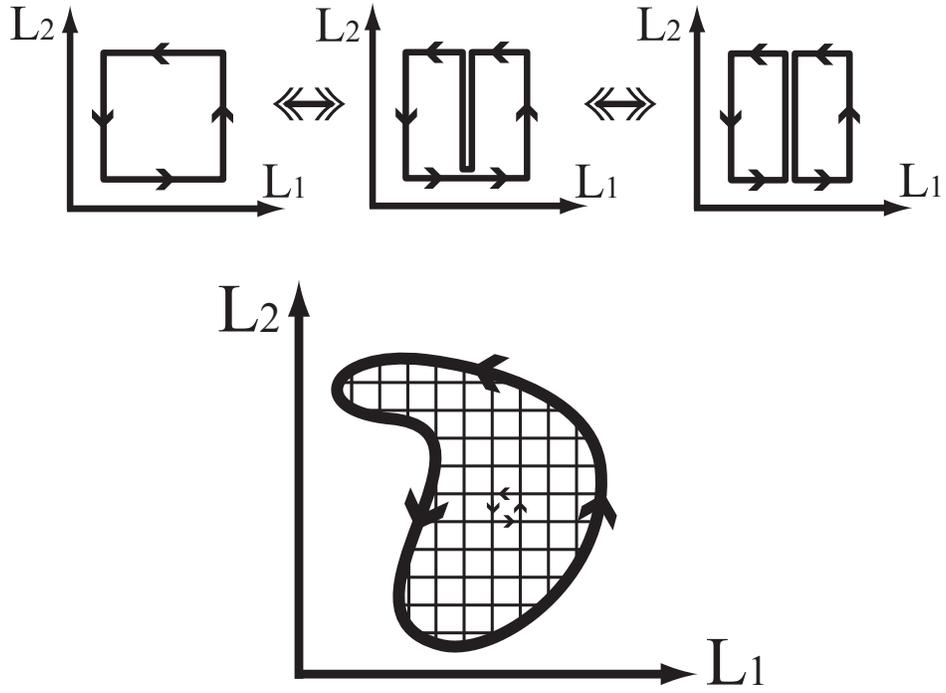}}
\caption{
The top three figures illustrate that since in Stokes flow no net migration is produced by any bead motions that merely reverse themselves,  the migration in any single cycle can be obtained as the sum of the migrations in two cycles into which the original cycle is divided.   Extending this principle,  the bottom figure illustrates that the migration in a cycle of arbitrary shape approaches that of the sum of migrations in small rectangular cycles into which the original cycle is divided.
}
\label{Fig-cycle-division}
\end{center}
\end{figure}

\begin{figure}
\begin{center}
\centerline{\includegraphics[width=5in]{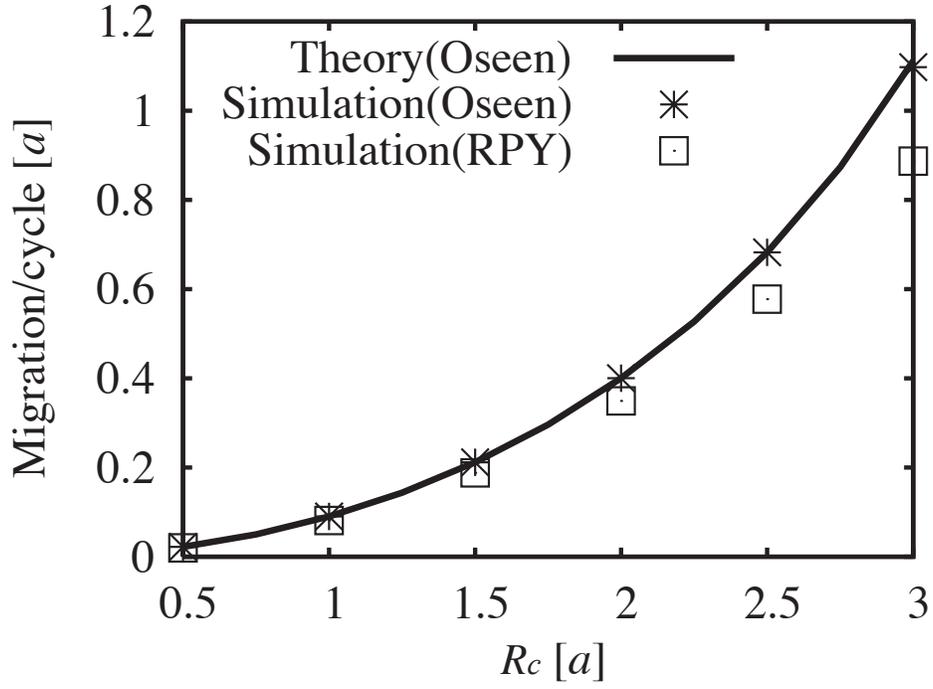}}
\caption{The migration per cycle for a CCW circular cycle centered at $(L_{1},L_{2})=(5a,5a)$ with radius $0.5a<R_{c}<3a$, obtained from Eq.~\ref{migration} (Theory), simulations with the Oseen tensor and the RPY tensor.
}
\label{Fig-migration}
\end{center}
\end{figure}

Now, we first calculate the migration for a small rectangular cycle, which we refer as $C_{r}$:
$(L_{1},L_{2})$$=$$(A,B)$$\to$$(A-\delta_{1},B)$$\to$$(A-\delta_{1},B-\delta_{2})$$\to$$(A,B-\delta_{2})$$\to$$(A,B)$.
The Oseen tensor is used for $\mathcal{H}_{ij}$ in Eq.~\ref{EqOfMotion} to calculate the migration ($L_{r}$) and we find:
\begin{equation}
\label{function-g}
	L_{r}=g(a,A,B)\delta_{1}\delta_{2},
\end{equation}
where $g(a,L_{1},L_{2})$ is a rational function (see Appendix A for the derivation), and $\delta_{1},\delta_{2}\ll a$.
The contour plot of the function $g$ is shown in Fig.~\ref{Fig-g}.
Note that $g(a,L_{1},L_{2})$ is always positive when the cycle is counter-clockwise (CCW) in $L_{1}$-$L_{2}$ space, and the migration along a clockwise (CW) cycle becomes $-L_{r}$.
Since a cycle that does not enclose area in $L_{1}$-$L_{2}$ space induces no migration per cycle (i.e. the Scallop Theorem~\cite{Purcell}), 
a cycle can be divided into two or more cycles 
whose migrations sum up to the migrations of the original cycle; see Fig.~\ref{Fig-cycle-division}.
Therefore, an arbitrary cycle can be divided into small rectangular cycles whose migrations sum to that of the original cycle,
and the migration per arbitrary cycle ($L_{c}$) can be written as:
\begin{equation}
\label{migration}
	L_{c}=\iint_\text{A} g(a,L_{1},L_{2})dL_{1}dL_{2},
\end{equation}
where $A$ is the area within the cycle.
The result of $L_{c}$ calculated using Eq.~\ref{migration} for a CCW circular cycle centered at $(L_{1},L_{2})=(5a,5a)$ with radius $0.5a<R_{c}<3a$ is plotted in Fig.~\ref{Fig-migration} along with our simulation results using the Oseen tensor and the RPY tensor.
The theoretical result from Eq.~\ref{migration} agrees well with the simulation results.
Note that the simulation results obtained using the RPY tensor deviates from our theoretical results based on the Oseen tensor when $R_{c}$ is large because the RPY tensor, which accounts for more accurate hydrodynamics, diverges from  the Oseen tensor when the distance between beads becomes comparable to the bead radius $a$.

\section{Dipolar-flow strength of three-bead swimmer}
\begin{figure}
\begin{center}
\centerline{\includegraphics[width=5in]{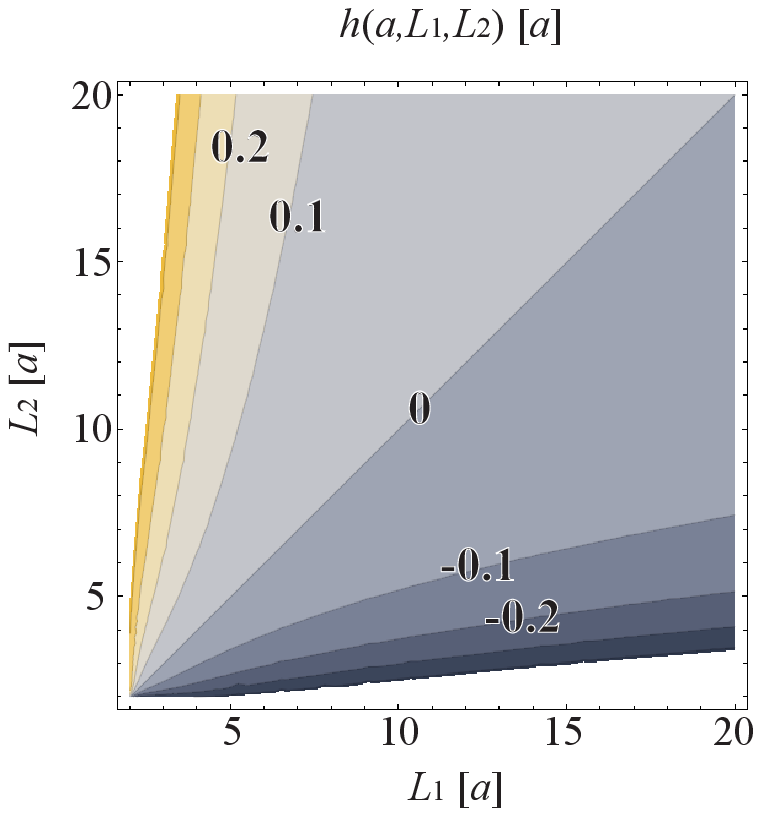}}
\caption{(color online). Contour plot of function $h(a,L_{1},L_{2})$ in Eq.~\ref{dipole}.
}
\label{Fig-h}
\end{center}
\end{figure}

\begin{figure}
\begin{center}
\centerline{\includegraphics[width=5in]{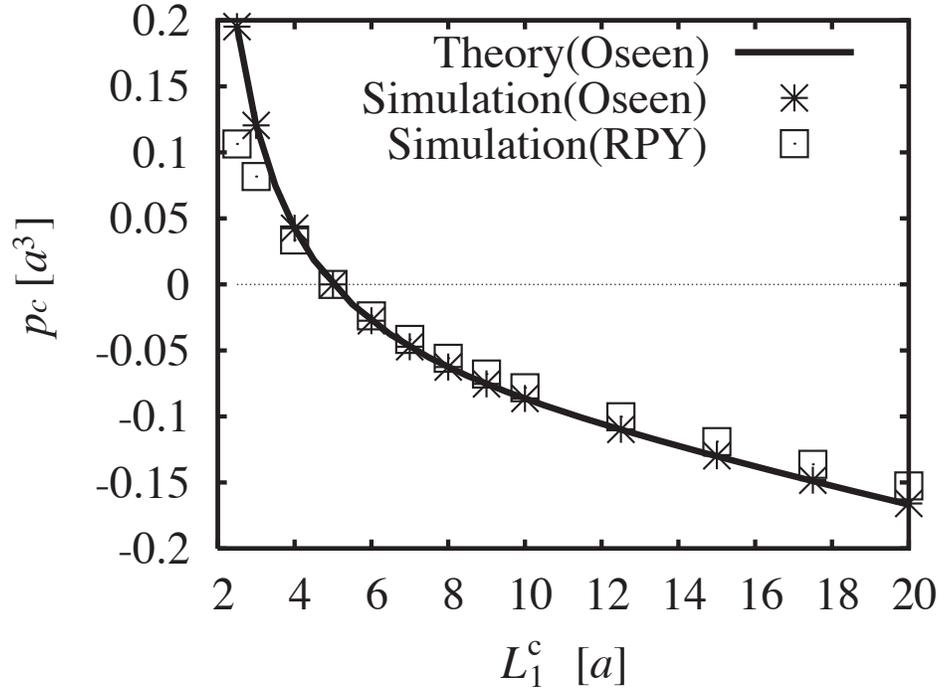}}
\caption{The strength of dipole flow, $p_{c}$, for a CCW circular cycle with radius $0.5a$ and centered at $(L_{1},L_{2})=(L_{1}^{c},5a)$ where $2a<L_{1}^{c}<20a$, obtained from Eq.~\ref{migration} (Theory), simulations with the Oseen tensor and the RPY tensor
}
\label{Fig-dipole}
\end{center}
\end{figure}

\begin{figure}
\begin{center}
\centerline{\includegraphics[width=5in]{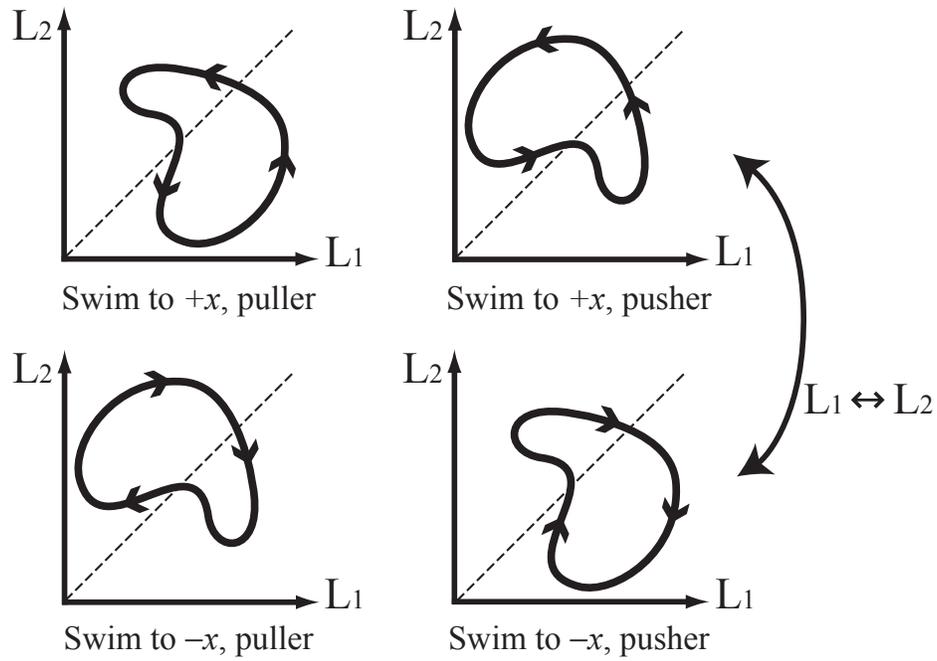}}
\caption{The change of the swimming behaviors by an inversion of the cycle-direction or an exchange of $L_{1}$ and $L_{2}$.
}
\label{Fig-puller-pusher}
\end{center}
\end{figure}

By taking advantage of the partitioning of a cycle described in the previous section,
we here evaluate the cycle-averaged flow field around a swimmer.
Once we find the analytical expression for the cycle-averaged flow for an arbitrary cycle, 
we can distinguish between puller and  pusher swimming from the  cycle path in $L_{1}$-$L_{2}$ space,
since its cycle-averaged far-field flow becomes dipolar as following:
\begin{equation}
\label{dipole-eq}
\langle \bm{v} (\bm{r})\rangle=\frac{p}{T} \left( 3[\hat{\bm{n}}\cdot\hat{\bm{r}}]^{2}-1 \right) \frac{\hat{\bm{r}}}{r^{2}}+\mathcal{O}(r^{-3}),
\end{equation}
where $\bm{r}$ is the position vector relative to the position of the center bead of a three-bead swimmer, $\hat{\bm{r}}$ is the unit vector of $\bm{r}$, $\hat{\bm{n}}$ is the unit vector pointing in the swimming direction, and $p$ represents the strength of dipole, where the flow is that of a pusher when $p>0$, that of a puller when $p<0$, and neither a pusher nor a puller when $p=0$.
Note that the far-field fluid displacement per cycle $\bm{d}(\bm{r})=T \langle \bm{v}(\bm{r})\rangle$ is an invariant of a cycle.
For a small CCW rectangular cycle $C_{r}$, the dipole strength $p_{r}$ can be evaluated analytically in a similar way as in the previous section using the Oseen tensor, assuming $\delta_{1}, \delta_{2}\ll a$ and $r\gg a$:
\begin{equation}
\label{function-h}
p_{r}=h(a,A,B)\delta_{1}\delta_{2},
\end{equation}
where $h(a,L_{1},L_{2})$ is a rational function (see Appendix A for the derivation).
Since a cycle enclosing no area in $L_{1}$-$L_{2}$ space induces no cycle-averaged flow field,
we can use the ``cycle-division'' to obtain the dipole strength induced by an arbitrary cycle:
\begin{equation}
\label{dipole}
	p_{c}=\iint_\text{A} h(a,L_{1},L_{2})dL_{1}dL_{2}.
\end{equation}
From the contour plot of the function $h(a,L_{1},L_{2})$ shown in Fig.~\ref{Fig-h}, we find that 
1) a cycle symmetric across the line $L_{1}=L_{2}$ induces no dipolar flow and the perturbation flow decays as $r^{-3}$, which is the same as a Najafi-Golestanian swimmer,
and 2) if a CCW cycle is entirely in the region $L_{1}>L_{2}$, the cycle-averaged flow is that of a puller, and if entirely in $L_{1}<L_{2}$, that of a pusher.
The result of $p_{c}$ calculated using Eq.~\ref{dipole} for a CCW circular cycle of radius $0.5a$ and centered at $(L_{1},L_{2})=(L_{1}^{c},5a)$ where $2a<L_{1}^{c}<20a$ is plotted in Fig.~\ref{Fig-dipole} along with our simulation results using the Oseen tensor and the RPY tensor.
The theoretical result agrees well with the simulation results except that the simulation results with the RPY tensor diverges from the theoretical result because of its higher order accuracy compared to the Oseen tensor.
Note also that if a CW cycle is entirely in the region $L_{1}>L_{2}$, the cycle-averaged flow is that of a pusher, and if entirely in $L_{1}<L_{2}$, that of a puller.
A cycle-direction inversion (CCW $\leftrightarrow$ CW) reverses the migration direction ($+x \leftrightarrow -x$), and the sign of the flow dipole ($p_{c}\leftrightarrow -p_{c}$) as shown in Fig.~\ref{Fig-puller-pusher}.
This is consistent with the fact that $p_{c}$ (or the function $h(a,L_{1},L_{2})$) changes sign upon a reversal of the cycle-direction.

\section{Rate of energy consumption of three-bead swimmer}
The rate of energy consumption during a cycle can be written as:
\begin{align}
e(s)&=f_{1}v_{1}+f_{2}v_{2}+f_{3}v_{3}\notag\\
&=f_{1}^{s}(v_{2}-v_{1})+f_{2}^{s}(v_{3}-v_{2})=f_{1}^{s}\dot{L}_{1}+f_{2}^{s}\dot{L}_{2},
\end{align}
where $s$ is the distance coordinate along a cycle in $L_{1}$-$L_{2}$ space.
Due to the linearity of the governing equation (i.e. the Stokes equation),
each spring force is proportional to the time derivative of the spring length, $f_{i}^{s}=c_{i}(s)\dot{L}_{i}$.
Using the chain rule of differentiation, we find:
\begin{align}
e(s)&=c_{1}(s)\dot{L}_{1}^{2}+c_{2}(s)\dot{L}_{2}^{2} \notag \\
&=c_{1}(s)\left( \frac{dL_{1}}{ds}\right)^{2}\left( \frac{ds}{dt}\right)^{2}+c_{2}(s)\left( \frac{dL_{2}}{ds}\right)^{2}\left( \frac{ds}{dt}\right)^{2}\notag \\
&=\left( \frac{ds}{dt}\right)^{2} \left[ c_{1}(s)\left( \frac{dL_{1}}{ds}\right)^{2}+c_{2}(s)\left( \frac{dL_{2}}{ds}\right)^{2}\right].
\end{align}
Therefore,
\begin{align}
&\frac{ds}{dt}=\sqrt{e(s)\left/\left[ c_{1}(s)\left( \frac{dL_{1}}{ds}\right)^{2}+c_{2}(s)\left( \frac{dL_{2}}{ds}\right)^{2}\right] \right.}, \label{ds-dt}\\
&dt=ds\sqrt{\left. \left[ c_{1}(s)\left( \frac{dL_{1}}{ds}\right)^{2}+c_{2}(s)\left( \frac{dL_{2}}{ds}\right)^{2}\right]\right/e(s)} \label{dt}.
\end{align}
By integrating Eq.\ref{dt} over a cycle,
we obtain the cycle time $T$,
\begin{equation}
\label{Cycle-Time}
T=\oint \sqrt{\left. \left[ c_{1}(s)\left( \frac{dL_{1}}{ds}\right)^{2}+c_{2}(s)\left( \frac{dL_{2}}{ds}\right)^{2}\right]\right/e(s)}~~ds.
\end{equation}
For a fixed cycle in $L_{1}$-$L_{2}$ space, $ds/dt$ and $T$ can be derived from Eq.~\ref{ds-dt} and \ref{Cycle-Time} once $e(s)$ is specified.
Since in the Stokes flow the rate of energy dissipation is quadratic in the velocity, It is easy to show  that the most energy efficient cycle motion $s_\text{eff}(t)$ along a cycle in a cycle time $T$ is achieved when the rate of energy consumption is constant through the cycle (see Appendix B for proof).

Assuming a constant rate of energy consumption $e$, the non-dimensional rate of energy consumption $\tilde{e}$ and the non-dimensional energy consumed per cycle $\tilde{E}$ can be defined as:
\begin{align}
\label{}
    & \tilde{e}=\frac{eT^{2}}{\eta a^{3}},  \\
    & \tilde{E}=\frac{ET}{\eta a^{3}},
\end{align}
where $E=eT$, and these non-dimensional parameters only depend on the cycle path in $L_{1}$-$L_{2}$ space.
Note that by setting $e$ to be a constant through a cycle, its value can be easily calculated from Eq.~\ref{Cycle-Time} with given cycle time $T$.

\section{Multi-dimensional swimmers}
The swimming direction of the one-dimensional three-bead swimmer is restricted to one dimension because of its linear geometry.
However, we can easily construct a two-dimensional swimmer by using beads that are not colinear (Fig.~\ref{Swimmers}).
Some micro-swimmers tumble and reorient the cell body by breaking symmetry of the swimming motion over the swimming axis.
For example, \textit{Chlamydomonas} swims straight with synchronous beating of two flagella and tumbles with asynchronous beating~\cite{Polin}, and \textit{E. coli} swims straight by rotating multiple helical flagella in a coherent rotational direction and tumbles by rotating at least one of the flagella in the reverse direction~\cite{Berg-book}.
We confirmed by simulations using the bead-spring model with time-dependent equilibrium spring lengths analogous to the one-dimensional swimmer that our two-dimensional swimmer also tumbles by breaking symmetry of the swimming motion as shown in the {cycle} history in Fig.~\ref{Swimmers}.
Therefore, by alternating between the runs and tumbles, this two-dimensional swimmer can swim in a two-dimensional random-walk trajectory.

We can also create a three-dimensional swimmer by using four or more non-coplanar beads, for example, on the vertexes of a tetrahedron or ``tetrumbbell''~\cite{tetrumbbell}.
This three-dimensional swimmer can swim straight with a periodic non-reciprocal motion analogous to that of the two-dimensional swimmer and tumble by breaking symmetry of the swimming motion (see Fig.~\ref{Swimmers}).
It is worth noting that a tetrumbbell with constant equilibrium bond-lengths but different spring constants can migrate in a shear flow in the vorticity direction at low Reynolds number~\cite{tetrumbbell}. 
This migration (or ``swimming'') also results from a periodic non-reciprocal deformation induced by the shear flow on the tetrumbbell. 

\begin{figure}
\begin{center}
\centerline{\includegraphics[width=1.8in]{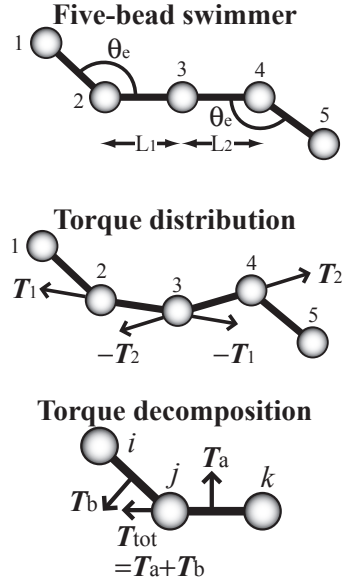}}
\caption{(Top) the configuration of the five-bead model. (Middle) the torque distribution on the swimmer.
(Bottom) a decomposition of a torque $\bm{T}_\text{tot}$ on three connected beads, which represents either $\bm{T}_{1}$, $\bm{T}_{2}$, $-\bm{T}_{1}$ or $-\bm{T}_{2}$ in the middle figure, into two torques $\bm{T}_{a}$ and $\bm{T}_{b}$ acting perpendicular to adjacent bonds.
}
\label{FiveBeadSwimmer}
\end{center}
\end{figure}

All of these models will be useful for studying the low Reynolds number hydrodynamics of micro-swimmers, the swimmer-swimmer interactions, and the collective dynamics of many swimmers.
To study the multi-swimmer interactions of ``corkscrew swimmers'' such as {\textit{E. coli}}, however, we need to include the effect of the angular velocity around the swimmer induced by the rotation of the flagellum and the counter-rotation of the cell-body.
In Fig.~\ref{FiveBeadSwimmer}, we depict a five-bead swimmer that induces a time-average flow field corresponding to either a pusher or a puller plus this angular velocity.
The five-bead swimmer swims via a periodic non-reciprocal motion of beads $2$, $3$ and $4$, just as the three-bead swimmer does, by changing the bond-lengths $L_{1}$ and $L_{2}$.
Additionally, torques $\bm{T}_{1}$ and $\bm{T}_{2}$, and counter-torques $-\bm{T}_{1}$ and $-\bm{T}_{2}$ are applied as shown in Fig.~\ref{FiveBeadSwimmer} to mimic the rotation of the flagellum and the counter-rotation of the cell-body.
Note that, for the total torque to be zero, the magnitudes of torques $\bm{T}_{1}$ and $\bm{T}_{2}$ must be the same, and also that when beads $2$, $3$ and $4$ are colinear, torques $-\bm{T}_{1}$ and $-\bm{T}_{2}$ cancel out.
Each of these torques is first decomposed into two torques perpendicular to the bonds of the swimmer, and then each of these decomposed into forces on beads.
For example, the force distribution on three connected beads $i$, $j$ and $k$ induced by a torque $\bm{T}_\text{tot}$, which represents either $\bm{T}_{1}$, $\bm{T}_{2}$, $-\bm{T}_{1}$ or $-\bm{T}_{2}$, is (see Fig.~\ref{FiveBeadSwimmer}),
\begin{align}
	&\bm{F}_{i}=\frac{T_{b}}{r_{ij}} \frac{\bm{T}_\text{tot}\times\bm{r}_{ij}}{\left| \bm{T}_\text{tot}\times\bm{r}_{ij} \right|},\\
	&\bm{F}_{j}=-\bm{F}_{i}-\bm{F}_{k},\\
	&\bm{F}_{k}=\frac{T_{a}}{r_{jk}} \frac{\bm{T}_\text{tot}\times\bm{r}_{ij}}{\left| \bm{T}_\text{tot}\times\bm{r}_{ij} \right|}.
\end{align}
To restrict the deformation of the swimmer, bending potentials are applied:
$\phi(\theta_{234}, 180^{\circ})$, $\phi(\theta_{123}, \theta_\text{e})$ and $\phi(\theta_{345}, \theta_\text{e})$,
where 
$\phi(\theta_{ijk}, \theta_\text{0})=\frac{1}{2} k_\text{b} (\cos \theta_{ijk} - \cos \theta_\text{0})^{2}$, $\theta_{ijk}$ is the angle formed by beads $i$, $j$, and $k$,
$\theta_0$ is the equilibrium bending angle,
$k_\text{b}$ is the bending potential constant,
and {we choose $\theta_\text{e}=160^{\circ}$ for our simulations}.
Although the bending potentials allow small deformations of the swimmer~\cite{Rigid-FiveBeadSwimmer}, neither the time-averaged flow field around the swimmer nor the swimming behavior are affected qualitatively for $k_\text{b}$ values as large or larger than that used here.
The time for $L_{1}$ and $L_{2}$ in a five-bead swimmer to traverse one side of a triangle in the configuration space is kept constant at the value $t_{0}$.

By solving Eq.~\ref{EqOfMotion} with the force distribution on beads obtained by adding together all forces associated with torques and all spring and bending forces, we obtain the time-averaged flow field around a five-bead pusher swimmer, which we find to be very similar to that obtained from a detailed bead-spring model (with 60 beads) of \textit{E. coli} with multiple helical flagella in a run~\cite{Bead-spring-Ecoli}.
Input parameters with physical units in this simulation are 
$R$ (maximum bond-length), $\eta$ (solvent viscosity), and $T$ (magnitude of torques).
Therefore, we scale length and time with $R$ and $\tau=\eta R^3/T$, respectively.
{We choose the simulation parameters $a$=$0.2R$, $\delta$=$0.5R$, $t_{0}$=$2.0\tau$, $H$=$1000T/R$, $k_\text{b}$=$2000T$, $\theta_\text{e}$=$160^{\circ}$ and $\Delta t$=$10^{-3}\tau$.}

\begin{figure}
\begin{center}
\centerline{\includegraphics[width=3in]{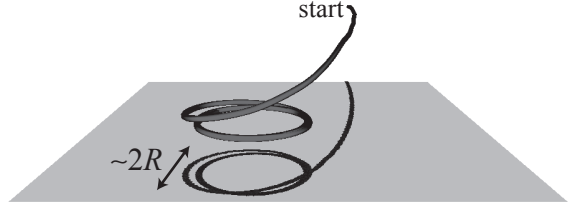}}
\caption{A trajectory of the five-bead swimmer near a wall. 
The trajectory is projected onto the wall to show the radius of the circular motion.
The swimmer is initially placed $2R$ away from and parallel to the wall. Simulation parameters are $a$=$0.2R$, $\delta$=$0.5R$, $t_{0}$=$2.0\tau$, $H$=$1000T/R$, $k_\text{b}$=$2000T$, $\theta_\text{e}$=$160^{\circ}$ and $\Delta t$=$10^{-3}\tau$.
}
\label{Trajectory}
\end{center}
\end{figure}

Simulations were also performed for the five-bead pusher near a wall by employing the RPY tensor with the wall effect included~\cite{Nazish} and a short-range repulsive potential between each bead and the wall with cut-off length $0.2R$.
We find that the swimmer is attracted to the wall and swims clockwise (see Fig.~\ref{Trajectory}) as observed in experiments and simulations with bacteria with helical flagella(um) (e.g. \textit{E. coli})~\cite{Frymier,Swimming-in-Circles,PhanThien}, despite the fact that the five-bead swimmer does not have a helical tail nor helical motion.
Therefore, the five-bead swimmer captures the experimentally observed behavior of \textit{E. coli} qualitatively, and will allow us to simulate the collective dynamics of micro-swimmers more realistically than before at modest cost.

We also confirmed that a three-dimensional tumbling motion of the five-bead swimmer can be induced by temporarily connecting beads $2$ and $4$ with a FENE-Fraenkel spring of constant equilibrium length $R$ while the bond-lengths $L_{1}$ and $L_{2}$ change in the same way as in a run, in other words, by forming a two-dimensional swimmer in a tumble with beads $2$, $3$ and $4$.
Simulations show that the trajectory of two five-bead pushers {in a run}, initially placed side by side and parallel to each other, agrees qualitatively with that obtained from a boundary element simulation of a modeled corkscrew swimmer, showing that two parallel swimmers first attract each other, and at the same time their swimming axes rotate in opposite directions~\cite{Ishikawa}.
We find that two side-by-side pullers behave similarly.

\section{Summary and future directions}

We have developed simple micro-swimmer models by extending the Najafi-Golestanian three-bead linear swimmer model to allow arbitrary history of bond-lengths $(L_{1},L_{2})$, by making the beads noncolinear to introduce tumbling, and by adding two additional non-collinear end beads to which torque and counter-torque are applied to induce a helical flow field around it, thus mimicking the swimming of flagellated bacteria.  
We have shown that both the migration distance and optimum energy consumption per cycle is an invariant of the cycle path, and can be computed, respectively, by area and path integrals over the cycle.  
For three-bead swimmers, we have also developed a simple criterion to distinguish puller from pusher swimmers, based on the cycle in $L_{1}$-$L_{2}$ space.  
Finally, we have shown that our five-bead corkscrew swimmer spirals towards flat surfaces and interacts hydrodynamically with other five-bead swimmers in ways qualitatively similar to that of much more refined models of \textit{E.coli}.
These minimal models of micro-swimmers will help in the study of the collective dynamics of micro-swimmers with a simplified but qualitatively accurate hydrodynamics.
These models can also be used to study the synchronization of multiple swimmers if the bond-lengths are changed by applying time-constant, equal and opposite forces on the beads at the ends of each bond until the bond-length reaches a designated length,
as proposed in a previous work~\cite{Putz}.

\begin{acknowledgments}
We acknowledge support from National Science Foundation (NSF) under grant NSEC EEC-0425626.
\end{acknowledgments}

\section*{APPENDIX A: ANALYTICAL EXPRESSIONS FOR THE MIGRATION AND THE DIPOLAR-FLOW STRENGTH INDUCED BY A SMALL RECTANGULAR CYCLE}
We analytically evaluate the migration per cycle induced by a small rectangular cycle $C_{r}$:
$(L_{1},L_{2})$$=$$(A,B)$$\to$$(A-\delta_{1},B)$$\to$$(A-\delta_{1},B-\delta_{2})$$\to$$(A,B-\delta_{2})$$\to$$(A,B)$.
Since the migration is independent of how the time is spent for strokes and determined only by the path in $L_{1}$-$L_{2}$ space because of the linearity of the governing equation, we assume the first stroke to be $(L_{1},L_{2})=(A-\frac{t}{T_{1}}\delta_{1},B)$ where $0<t<T_{1}$.
By solving the following simultaneous equations derived from Eq.~\ref{EqOfMotion} with the Oseen tensor, we obtain $(f_{1}^{s}(t),f_{2}^{s}(t))$,
\begin{align}
\label{}
    &\dot{L}_{1}=v_{2}-v_{1}=	\frac{2f_{1}^{s}-f_{2}^{s}}{6\pi\eta a}+
    											\frac{-2f_{1}^{s}+f_{2}^{s}}{4\pi\eta L_{1}}+
												\frac{f_{2}^{s}}{4\pi\eta L_{2}}+
												\frac{-f_{2}^{s}}{4\pi\eta (L_{1}+L_{2})},   \\
    &\dot{L}_{2}=v_{3}-v_{2}=	\frac{2f_{2}^{s}-f_{1}^{s}}{6\pi\eta a}+
    											\frac{-2f_{2}^{s}+f_{1}^{s}}{4\pi\eta L_{2}}+
												\frac{f_{1}^{s}}{4\pi\eta L_{1}}+
												\frac{-f_{1}^{s}}{4\pi\eta (L_{1}+L_{2})}.
\end{align}
Then, we can calculate $(v_{1}(t),v_{2}(t),v_{3}(t))$ using $(f_{1}^{s}(t),f_{2}^{s}(t))$,
\begin{align}
\label{}
    &v_{1}=	\frac{-f_{1}^{s}}{6\pi\eta a}+
    				\frac{f_{1}^{s}-f_{2}^{s}}{4\pi\eta L_{1}}+
					\frac{f_{2}^{s}}{4\pi\eta (L_{1}+L_{2})},\\
    &v_{2}=	\frac{f_{1}^{s}-f_{2}^{s}}{6\pi\eta a}+
    				\frac{-f_{1}^{s}}{4\pi\eta L_{1}}+
					\frac{f_{2}^{s}}{4\pi\eta L_{2}},\\
   &v_{3}=	\frac{f_{2}^{s}}{6\pi\eta a}+
    				\frac{f_{1}^{s}-f_{2}^{s}}{4\pi\eta L_{2}}+
					\frac{-f_{1}^{s}}{4\pi\eta (L_{1}+L_{2})}.
\end{align}
Finally, we find the center-of-mass migration velocity $v_\text{mig}(t)=(v_{1}+v_{2}+v_{3})/3$ to be
\begin{multline}
\label{v-mig}
v_\text{mig}(t)=\frac{a}{3}[-2 L_1 L_2 (L_1+L_2) \{(2 {\dot{L}_1}+{\dot{L}_2}) L_1^2-({\dot{L}_1}+2 {\dot{L}_2}) L_2^2\}+\\
3 a \{(2 {\dot{L}_1}+{\dot{L}_2}) L_1^4+(2 {\dot{L}_1}+{\dot{L}_2}) L_1^3 L_2+(-{\dot{L}_1}+{\dot{L}_2}) L_1^2 L_2^2-({\dot{L}_1}+2 {\dot{L}_2}) L_1 L_2^3-({\dot{L}_1}+2 {\dot{L}_2}) L_2^4\}]/\\
\{ -4 L_1^2 L_2^2 (L_1+L_2)^2+4 a L_1 L_2 (L_1+L_2) (L_1^2+3 L_1 L_2+L_2^2)+
3 a^2 (L_1^4-2 L_1^3 L_2-5 L_1^2 {L_2}^2-2 {L_1} {L_2}^3+{L_2}^4)\}.
\end{multline}
Note that this solution for the migration velocity is different from that in Ref.~\cite{PRE-Golestanian} because we do not use approximation $a/L_{i}\ll1$, which is used in their calculation.

We find the migration induced by the first stroke using Eq.~\ref{v-mig},
\begin{multline}
\label{ }
\int_{0}^{T_{1}}v_\text{mig}(t)dt=
\frac{a \delta_{1}}{3} \{-2 A B (-2 A^3-2 A^2 B+A B^2+B^3)+3 a (-2 A^4-2 A^3 B+A^2 B^2+A B^3+B^4)\} /\\
 \{-4 A^2 B^2 (A+B)^2+4 a A B (A^3+4 A^2 B+4 A B^2+B^3)+3 a^2 (A^4-2 A^3 B-5 A^2 B^2-2 A B^3+B^4)\}+\mathcal{O}(\delta_{1}^{2}).
\end{multline}
By adding up the migrations induced by the other three strokes in the cycle, which can be calculated in the same way, we find the total migration for this rectangular cycle to be $L_{r}=g(a,A,B)\delta_{1}\delta_{2}$, where $\delta_{1},\delta_{2}\ll a$ and $g(a,A,B)$ is
\begin{multline}
g(a,A,B)=\frac{1}{3} a \{16 A^2 B^2 (A+B)^2 (A^4+2 A^3 B+A^2 B^2+2 A B^3+B^4)+\\
36 a^2 (A^2+3 A B+B^2) (A^6+3 A^5 B+3 A^4 B^2+A^3 B^3+3 A^2 B^4+3 A B^5+B^6)-\\
27 a^3 (A+B) (2 A^6+5 A^5 B+6 A^4 B^2-5 A^3 B^3+6 A^2 B^4+5 A B^5+2 B^6)-\\
12 a A B (A+B) (4 A^6+14 A^5 B+17 A^4 B^2+10 A^3 B^3+17 A^2 B^4+14 A B^5+4 B^6)\}/\\
\{-4 A^2 B^2 (A+B)^2+4 a A B (A+B) (A^2+3 A B+B^2)+3 a^2 (A^4-2 A^3 B-5 A^2 B^2-2 A B^3+B^4)\}^2.
\end{multline}

The strength of the dipolar-flow induced by the same cycle $C_{r}$ can be evaluated in a similar way by calculating the cycle-averaged flow:
\begin{equation}
\label{Cycle-ave-flow}
\langle \bm{v}(\bm{r})\rangle=\frac{1}{T}\int_{0}^{T}\sum_{i=1}^{3}\mathcal{H}_{ij}(\bm{r}-\bm{r}_{i}(t))\cdot \bm{f}_{i}(t)dt
\end{equation}
where $\bm{r}$ is the position vector originated at the position of the center bead of a three-bead swimmer, and $\bm{r}_{i}(t)$ is obtained by 
$\bm{r}_{i}(t)=\bm{r}_{i}(0)+\int_{0}^{t}\bm{v}_{i}(t)dt$.
We find the strength of dipolar flow to be $p_{r}=h(a,A,B)\delta_{1}\delta_{2}$ by comparing the result of Eq.~\ref{Cycle-ave-flow} with Eq.~\ref{dipole-eq} under conditions $\delta_{1},\delta_{2}\ll a$ and $r\gg a$, where $h(a,A,B)$ is
\begin{multline}
h(a,A,B)=\frac{1}{2}a^2 (-A+B) \{-12 a A B (A+B) (2 A+B) (A+2 B) (A^2+A B+B^2 )^2+\\
9 a^2 (A^2+3 A B+B^2 ) (2 A^3+3 A^2 B+A B^2+B^3 ) (A^3+A^2 B+3 A B^2+2 B^3 )+\\
4 A^2 B^2 (A+B)^2  (2 A^4+7 A^3 B+11 A^2 B^2+7 A B^3+2 B^4 ) \} /\\
\{-4 A^2 B^2 (A+B)^2+4 a A B (A+B)  (A^2+3 A B+B^2 )+3 a^2  (A^4-2 A^3 B-5 A^2 B^2-2 A B^3+B^4 ) \}^2.
\end{multline}

\section*{APPENDIX B: MOST ENERGY EFFICIENT CYCLE MOTION OF THREE-BEAD SWIMMER}
Here, we show that the most energy efficient cycle motion $s_\text{eff}(t)$ along a cycle in a cycle time $T$ is achieved when the  rate of energy consumption is constant through the cycle.
We show this by evaluating the change of the total energy consumed in a cycle when a portion of a cycle-motion is sped up and another is slowed down, while keeping the cycle time $T$ constant.
Let us change the time spent for two small line segments on a cycle, $\Delta s_{p}$ and $\Delta s_{q}$, which is passed through during a small time interval $\Delta T$, keeping the segment lengths constant.
\begin{align}
\label{}
   &  \Delta T \to (1+\epsilon)\Delta T \qquad\text{in the segment }\Delta s_{p}.\\
   &  \Delta T \to (1-\epsilon)\Delta T \qquad\text{in the segment }\Delta s_{q}.
\end{align}
As a result, we find the following changes:
\begin{align}
\label{}
   &  \frac{\Delta s_{p}}{\Delta T} \to \frac{1}{1+\epsilon} \frac{\Delta s_{p}}{\Delta T},~~
   v_{i}\to\frac{v_{i}}{1+\epsilon},~~
   f_{i}^{s}\to\frac{f_{i}^{s}}{1+\epsilon}\qquad\text{in the segment }\Delta s_{p},\\
   &  \frac{\Delta s_{q}}{\Delta T} \to \frac{1}{1-\epsilon} \frac{\Delta s_{q}}{\Delta T},~~
   v_{i}\to\frac{v_{i}}{1-\epsilon},~~
   f_{i}^{s}\to\frac{f_{i}^{s}}{1-\epsilon}\qquad\text{in the segment }\Delta s_{q}.
\end{align}
These changes cause the changes in the rate of energy consumption,
\begin{align}
\label{}
   &  e_{p}=\sum_{i=1}^{3}f_{i}v_{i} \to \frac{e_{p}}{(1+\epsilon)^{2}}\qquad\text{in the segment }\Delta s_{p},\\
   &  e_{q}=\sum_{i=1}^{3}f_{i}v_{i} \to \frac{e_{q}}{(1-\epsilon)^{2}}  \qquad\text{in the segment }\Delta s_{q}.
\end{align}
The energy consumed in these segments becomes,
\begin{align}
\label{}
   &  e_{p}\Delta T \to \frac{e_{p}}{(1+\epsilon)^{2}}\cdot(1+\epsilon)\Delta T= \frac{e_{p}}{(1+\epsilon)}\Delta T\qquad\text{in the segment }\Delta s_{p},\\
   &  e_{q}\Delta T \to \frac{e_{q}}{(1-\epsilon)^{2}}\cdot(1-\epsilon)\Delta T= \frac{e_{q}}{(1-\epsilon)}\Delta T\qquad\text{in the segment }\Delta s_{q},
\end{align}
where $e_{p}$ and $e_{q}$ are the rate of energy consumption before the change of time interval in segments $\Delta s_{p}$ and $\Delta s_{q}$, respectively.
If we assume {that the original rate of energy consumption are constant}, or $e_{p}=e_{q}=e$, the change of the total energy consumed in a cycle after the change of time intervals is
\begin{equation}
\label{ }
\Delta E=\left\{ \frac{e\Delta T}{1+\epsilon} + \frac{e\Delta T}{1-\epsilon}\right\}-2e\Delta T= 2 \frac{\epsilon^{2}}{1-\epsilon^{2}}e\Delta T>0.
\end{equation}
Therefore, if the rate of energy consumption changes in the course of a cycle while keeping $T$ constant, the energy consumed in a cycle increases.
In other words, the most energy efficient cycle motion along a cycle in cycle time $T$ is achieved when the rate of energy consumption is constant through the cycle.


\clearpage

\clearpage

\clearpage

\clearpage

\clearpage

\clearpage

\clearpage

\clearpage

\clearpage

\end{document}